\documentstyle[preprint,prb,aps,epsf]{revtex}
\begin{document}
\draft
\title{An elastic model for the In--In correlations in In$_{x}$Ga$_{1-x}$As semiconductor alloys}
\author{A. S. Martins,}
\author{Belita Koiller and R. B. Capaz}
\address{Instituto de F\'{\i}sica, Universidade Federal do Rio de
Janeiro}
\address{Cx. Postal 68.528, Rio de Janeiro, 21945-970, Brazil}
\date{\today}
\maketitle

\begin{abstract}
Deviations from randomicity in In$_{x}$Ga$_{1-x}$As semiconductor alloys induced by elastic effects are investigated within the Keating potential.
Our model is based on Monte Carlo simulations on large (4096 atoms) supercells, performed 
with two types of boundary conditions: Fully periodic boundary conditions represent the 
bulk, while periodic boundary conditions along the $x$ and $y$ directions and a free surface in the $z$ direction simulate the epitaxial growth environment.
We show that In--In correlations identified in the bulk tend to be enhanced 
in the epitaxially grown samples. 
\end{abstract}

\pagebreak

\section{Introduction} 

In recent years, order versus disorder in substitutional semiconductor
alloys has motivated several theoretical \cite{rev,Teo1,Teo2,Teo3,Teo4} and experimental \cite{zheng,Chao,Pfister,Holo,Salemink} studies.
Special emphasis is given to the understanding of the physical mechanisms behind observed deviations from random distributions in the atomic positions. The distribution of different species over the atomic sites in such compounds is largely responsible for variations in electronic properties such as the band gap, the density of states at the Fermi energy and electron confinement levels. 
Moreover, the atomic correlations in these systems are responsible for processes such as interface segregation and clustering, which are fundamental in defining the atomic scale structure and roughness of heterostructure interfaces. 

In particular, In$_{x}$Ga$_{1-x}$As semiconductor alloys with $x\leq 20\%$ 
have been the subject of experimental studies indicating deviations from randomicity in their atomic configurations \cite{zheng,Chao,Pfister}. 
Zheng {\it et al.} \cite{zheng} identified a strong correlation between In atoms in $x=20\%$ samples. 
Clusters of 2 -- 3 In atoms were reported to form preferentially along the [001] growth direction, in second-nearest-neighbor (2nn) positions. 
Chao {\it et al.} \cite{Chao} investigated  $x=5\%$ samples and, in apparent contradiction with Zheng's results, no strong correlations were found in for 2nn pairs along the growth direction, whereas important anti-correlation for first-nearest-neighbor (1nn) pairs along [110] was reported \cite{footnote}. 
The absence of clustering of In atoms in $ x=12\% $ quantum wires was 
also noted by Pfister {\it et al.} \cite{Pfister} . 
Chao and Zheng emphasize that the mechanism behind the clustering/anticlustering effects in these alloys is the strain induced by In atoms, due to the $7\%$ mismatch between the GaAs and InAs lattice parameters. Both studies \cite{zheng,Chao} were based on In$_{x}$Ga$_{1-x}$As alloys  grown using molecular beam epitaxy (MBE) on GaAs (001) substrates at temperatures $\sim 800K$.
The alloy samples were kept below the critical thickness on GaAs substrate.
Cross-sectional images for cleaved $(110)$ \cite{zheng} and $(1\overline{1}0$ \cite{Chao} surfaces, obtained from scanning tunneling microscopy (STM) were used to determine the In distribution in the alloy. We follow here the orientation convention 
of Pashley {\it et al.}  \cite{Pashley}. 

The only apparent reason for the discrepancy concerning the second-neighbor clustering tendency in these experiments is the In concentration range: Cluster formation was reported in $x=20\%$ samples \cite{zheng}, while no second-neighbor correlation was identified in samples with smaller values of $x$ \cite{Chao,Pfister}. We investigate this possibility presenting a theoretical study for In-In pair correlations within a Keating model potential \cite{keating} to describe the elastic interactions in the alloy. The Monte Carlo Metropolis algorithm is used to determine the thermodynamic equilibrium configuration at finite temperatures. Phase-diagram calculations predict that InGaAs alloys are completely miscible at the growth temperature \cite{Teo2}, but no detailed theoretical analysis of pair correlation functions has been found in the literature. Therefore, one of the goals of the present work is to determine if the resulting In-In correlations are a consequence of bulk thermodynamics or if growth kinetics is important. Our simulations are then performed within two models, according to the boundary conditions: A bulk model (fully periodic) and an epitaxial growth model (periodic only in the $x$ and $y$ directions). 

\section{Formalism} 

Following Ref.~\cite{Chao}, we define the pair correlation function, $C\left( {\bf r}_{12}\right) $  as 
\begin{equation}
C\left( {\bf r}_{12}\right) =  \frac{P\left( {\bf r}_{12}\right) }
{R\left( {\bf r}_{12}\right) },
\end{equation}
where $P({\bf r}_{12}) = P\left( {\bf r}_{1},{\bf r}_{2}\right) $ is the 
probability of finding two In atoms at ${\bf r}_{1}$ and ${\bf r}_{2}$, 
which, for homogeneous alloys, is only a function of the 
relative position ${\bf r}_{12} = {\bf r}_{1}-{\bf r}_{2}$,  and 
$R( {\bf r}_{12}) $ is the equivalent quantity calculated for a random alloy 
with the same composition. 
If $C\left( {\bf r}_{12}\right) >1,$ we have  correlation
between the atoms that form the pair, and if 
$C\left( {\bf r}_{12}\right) <1,$ we have anticorrelation. Correlation indicates an effective attraction between the atoms, and anticorrelation implies a net repulsion. 

The Keating potential provides a good description for elastic energies in III-V semiconductor alloys \cite{JLmartins,Mattias}. 
It is given by  
\begin{equation}
U=\sum \limits_{ij}\frac{3}{8d_{ij}^{2}}\alpha _{ij}\left[ {\bf r}%
_{ij}^{2}-d_{ij}^{2}\right] ^{2}+\sum \limits_{\angle ijk}\frac{3}{%
8d_{ij}d_{jk}}\beta _{ijk}\left[ {\bf r}_{ij}\cdot {\bf r}_{jk}+\frac{%
d_{ij}d_{jk}}{3}\right] ^{2},
\end{equation}
where $d_{ij}$ is the equilibrium bond length between atoms $i$ and $j$
in the corresponding zincblend binary compound, ${\bf r}_{ij}$ is the relative position vector between nearest neighbors in 
different sublattices and $\alpha _{ij}$ 
and $\beta_{ijk}$ are the bond-stretching and bond-bending constants, respectively.
The elastic energy thus results from two contributions: 
The first summation (performed over nearest-neighbor pairs) refers to the 
excess energy due to the bond-length variations and
the second term (three-center term) gives the contribution due to the deviations of bond angles from their ideal tetrahedral values. 
Values for $d_{ij},$ $\alpha _{ij}$ and $\beta _{ijk}$ 
were taken from Ref.~\cite{JLmartins}. 
In particular, the difference between $d_{\rm GaAs}=2.448\,$\AA~and 
$d_{\rm InAs}=2.622$\,\AA~leads to the structural mismatch in the system.
For heterogeneous bond angles, $\beta _{ijk}$ was taken as the geometric 
mean of the bond-bending constants for the binary compounds: 
\begin{equation}
\beta _{ijk}=\sqrt{\beta _{iji}\beta _{kjk}}.
\end{equation}

Calculations are carried out in supercells with $N_{x}N_{y}N_{z}$
conventional cubic unit cells of the fcc lattice along the $x$, $y$ and $z$ directions, respectively, resulting in a system with $N=8N_{x}N_{y}N_{z}$ atoms. 
In$_{x}$Ga$_{1-x}$As semiconductor alloys are modeled by supercells in which ${x\cdot N}/2$ sites of the group-III sublattice are occupied by In atoms and the remaining sites by Ga, while the $N/2$ group-V sublattice sites are occupied by As atoms. For a given configuration, the energy is minimized by relaxing all the atomic positions in the supercell. Initially, we set the atoms to be randomly distributed, and evolution to thermodynamic equilibrium at the growth temperature ($T=800$ K) is carried out through the Monte Carlo Metropolis algorithm, via first-neighbor In~$\leftrightarrow$~Ga positions exchanges. The calculations are performed at constant volume. In the bulk model (Section 3) we use the Vegard's law \cite{veg} to give the approximated value for the lattice parameter and for the epitaxial growth model (Section 4) we keep the lattice parameter to the GaAs value, aiming to describe a strained growth process.

\section{Bulk model}
For the bulk model, after $\sim 10^3$ MC steps the system attains thermodynamic equilibrium and statistically independent ($\sim 500$ MC steps apart) alloy configurations are collected into an ensemble to calculate the pair correlation function.
This procedure is repeated for 6 initial random alloy configurations, leading to 
analyzed ensembles of at least 100 configurations for each temperature and composition.

The third column of Table \ref{bulk} shows our results for $x=0.05$ and $x=0.20$ 
in In$_{x}$Ga$_{1-x}$As. Notice an anticorrelation (repulsion) between 1nn pairs and a correlation (attraction) between 2nn pairs. These results are in qualitative agreement with experiments. However, quantitative agreement is poor: Chao's \cite{Chao} experiments indicate a much stronger anticorrelation for 1nn pairs than what is predicted by bulk thermodynamics. In addition, the observed experimental discrepancy regarding the 2nn correlation cannot be explained as a composition effect within the bulk model: There is no increase in the calculated correlation upon increasing In concentration from 5\% to 20\%. 

\begin{figure}
\setlength{\unitlength}{1mm}
\begin{picture}(150,150)(0,0)
\put(-10,-50){\epsfxsize=16cm\epsfbox{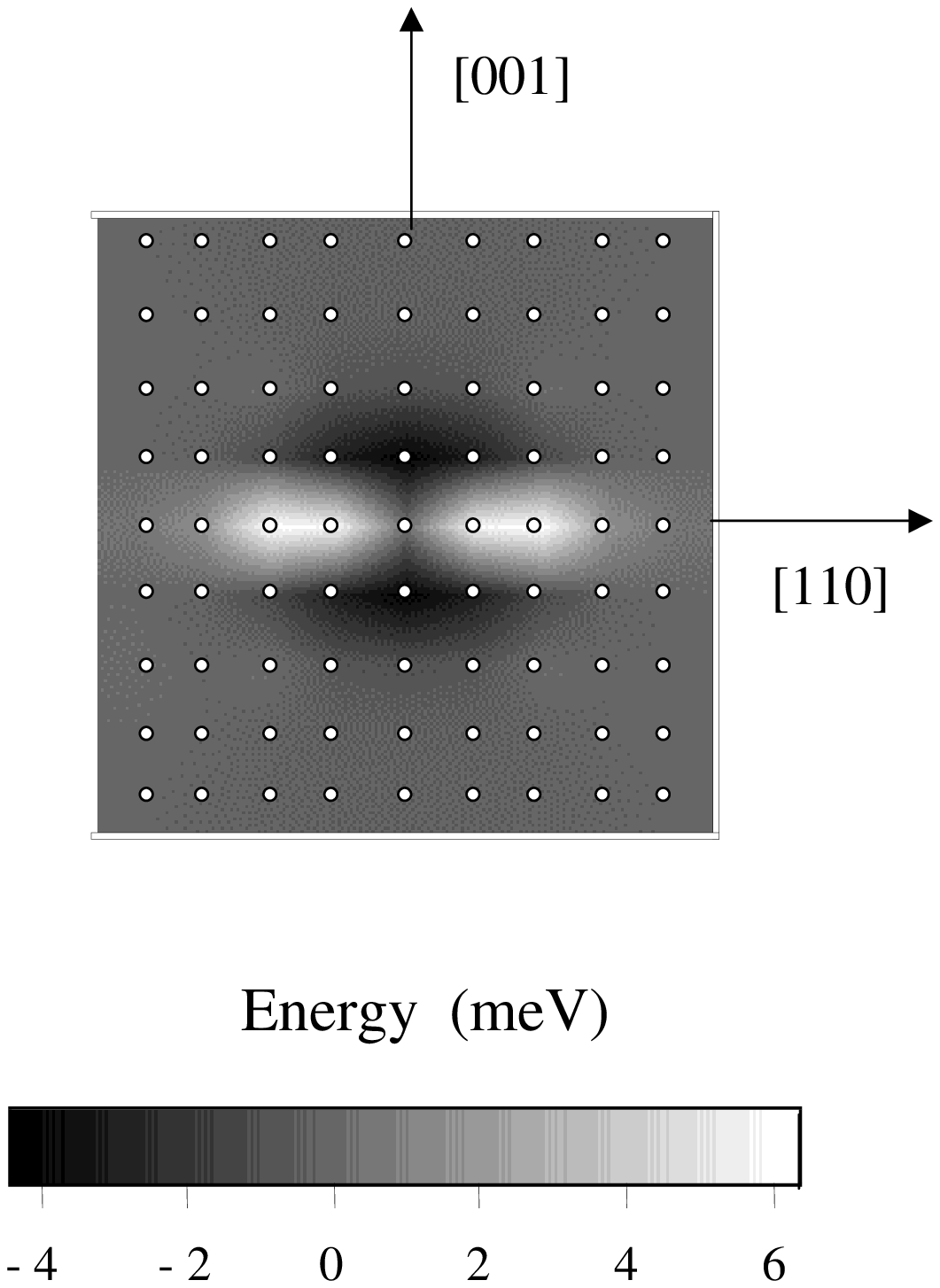}}
\end{picture}
\caption{In-In elastic energy distribution in the (1$\overline{1}$0) plane. 
Dark (light) regions correspond to lower (higher) energies.
Open circles indicate the atomic sites in the group-III sublattice. 
In this plane, the $[001]$ direction favors In pair formation and the $[110]$  direction does not.}
\label{energy} 
\end{figure}

Although not significantly larger than one, the 2nn correlation is clearly present. It is somewhat surprising that there can be such an effective attraction (for 2nn pairs) between In atoms based on elastic effects alone. Since In atoms are larger than Ga ones, one should expect that In impurities should compress the lattice locally and therefore repel each other. To see this in more detail, let's consider the interaction of two isolated In impurities in GaAs. We define a pair interaction energy as 
\begin{equation}
\Delta E=U-2E_{0},
\end{equation}
where $U$ is the Keating energy for the supercell with two impurities and $E_{0}$ the elastic energy for a single impurity, calculated to be $E_{0}=138$ meV. The interaction energy is displayed in Table \ref{tab}. Notice a net attraction (negative energy) along the [001] direction and a net repulsion (positive energy) along [110], consistent with the MC results. 
The interaction energy is therefore highly anisotropic, as can also be visualized in Figure \ref{energy}. 
These results can in fact be explained by the intricate geometry of the zincblende structure, as we can see by considering the mean Ga-As bond length deviations induced at nearby site by a single In impurity  
\begin{equation}
{\Delta L_i}= \frac{1}{4} \sum\limits_{j=1}^4|{\bf r}_{ij}|- d_{\rm GaAs}~,
\end{equation}
where $i$ is a Ga site and the sum in $j$ is performed over the As sites nearest neighbors to $i$. Negative (positive) $\Delta L$ at a given site means that the lattice is locally compressed (expanded) at that site by the nearby In impurity, and therefore suggests that putting a second In impurity at that site will be energetically unfavorable (favorable). One can see from Table \ref{tab} that even a larger impurity such as In in GaAs can produce a local {\it expansion} of the zincblende structure along certain directions (for instance, the [001] direction). 
An exact opposition between the signs of $\Delta L$ and  $\Delta E$ can also be readily seen ion the Table, thus supporting the argument given above. The highly anisotropic nature of the pair interactions is in agreement with first-principle calculations for these energies \cite{Teo2}. Interaction energies are determined there from a series expansion in terms of multiatom contributions, fitted to a set of ordered structures. Their results are in qualitative agreement with the present Keating model calculations, giving the most repulsive interactions between first and fourth neighbors [(110) and (220)], while the less repulsive ones correspond to second and third neighbors [(002) and (211) respectively]. 
Quantitatively, our results in Table \ref{tab} are lower than those in Ref.\cite{Teo2} by several meV and, in particular, no attractive pair interactions result from the 
first-principle fits.

\section{The epitaxial growth model}

Aiming at a more realistic model describing the experiments, 
we propose a simple description of epitaxial growth: 
{\it (i)} No periodic boundary conditions are imposed in the $z$ (growth) direction; 
{\it (ii)} The first two (bottom) monolayers, with $N_x=N_y=8$, represent the GaAs substrate, and the atomic species are fixed as in bulk GaAs. Atomic positions are fixed according to GaAs only in the $z=0$ plane; 
{\it (iii)} Growth in the $z$ direction proceeds by adding a single alloy monolayer
at a time, until $N_z=8$: 
The In~$\leftrightarrow$~Ga exchanges in the MC algorithm are restricted to the most 
recently grown monolayer (free surface) and the next monolayer is introduced only after thermal equilibrium has been attained. Atomic correlations between successive monolayers are thus induced by the In distribution in the inner layers. This is a reasonable assumption since, at typical MBE growth rates, the time scales for atomic processes at the free surface are much shorter than those in the previously grown planes. In addition, as can be seen in Table \ref{segreg}, there is an elastic-induced attraction between In impurities and the surface. Allowing bulk-surface In exchanges would produce In segregation at the growing surface, an effect that is known to occur but which we do not intend to describe here. In this present model, we do not treat dangling bonds, i.e., the surface is described simply as a truncated semi-infinite solid.
 
In the fourth column of Table \ref{bulk}, we present the pair correlation function for 5\% and 20\% In concentrations in In$_{x}$Ga$_{1-x}$As. There are now two inequivalent in-plane 1nn pairs directions, $[110]$ and $[1\overline{1}0]$. 
For the epitaxial growth model, the configurational averages were performed over 20 different configurations for each concentration and temperature. 
The statistical ensemble in the epitaxial growth model is considerably 
smaller than for the bulk model, since each growth process here contributes with a single configuration to the ensemble. 
Moreover, the bottom monolayers are fixed in composition to be of GaAs, which decreases the total number of In pairs in each generated configuration.
This leads correspondingly to larger statistical error bars.

We note that for $x=5\%$, $C\left( {\bf r}_{12}\right)$ 
along  $[110]$ is significantly reduced with respect to the bulk model behavior 
(see Table \ref{bulk}), while pairs along $[1\overline{1}0]$ become essentially uncorrelated. 
This is an interesting feature of the epitaxial growth model: 
The free surface induces important anisotropy between the $[110]$ and $[1\overline{1}0]$  
directions, and the correlation function in the bulk is approximately the
average between those calculated for the two inequivalent directions in the epitaxial 
growth model.
The experimental results of Ref.\cite{Chao} give $C({\bf r_{12}})=0.25$ 
along $[110]$ for $x=5\%$ and $T=800$ K. 
Our corresponding values for the bulk and epitaxial models, $C({\bf r_{12}})=0.83$ 
and 0.56 respectively, indicate that epitaxial growth is a decisive ingredient in the observed experimental behavior.
Although the agreement is still not quantitative, our results confirm that the observed 1nn anti-correlation has an elastic origin. For 2nn pairs with $x=20\%$, we do not observe significant increase of the correlation function above the random value. However, the correlation enhancement in going from 5\% to 20\% seems to be present in the epitaxial growth model, although error bars are too large to allow a more precise statement. Therefore, only elastic effects and a simple epitaxial growth do not provide a clear and quantitative explanation for the clusters formation reported in Ref.\cite{zheng}, even though we detect a {\it tendency} for In-In attraction in the growth direction in many of the cases studied. Perhaps a more sophisticated growth model, including aspects such as surface reconstruction and step-mediated growth, is necessary to better evaluate such results.

{\bf Acknowledgments} This work was partially supported by CNPq, CAPES, FAPERJ and FUJB (Brazil).

\vfill\eject

\eject

\begin{table}
\caption{\label{bulk}} Calculated (bulk and epitaxial growth models) and experimental correlation functions calculated for different In concentrations ($x$) in the alloy and In-In relative positions ({\bf r}) {\em in an undistorted {\rm GaAs} supercell} in units of half of the conventional GaAs lattice parameter $a$;. 

\bigskip

\begin{tabular}{|c|c|c|c|c|}
\hline
{\it x} & {\bf r} $\left(\times \frac{a}{2}\right)$ & Bulk & Epitaxial growth & Experiment  \\ \hline
0.05	& $(110)$ & $0.83 \pm 0.03$ & $0.56 \pm 0.11$ & 0.25 \cite{Chao} \\ 
	& $(1\overline{1}0)$ & $0.83 \pm 0.03$ & $0.93 \pm 0.09$ &  \\
      & $(002)$ & $1.10 \pm 0.05$ & $0.89 \pm 0.12$ & \\ \hline
0.20	& $(110)$ & $0.84 \pm 0.01$ & $0.86 \pm 0.03$ &  \\ 
	& $(1\overline{1}0)$ & $0.84 \pm 0.01$ & $0.96 \pm 0.02$ &  \\
      & $(002)$ & $1.07 \pm 0.01$ & $1.03 \pm 0.03$ & $>1$ \cite{zheng}\\ \hline

\end{tabular}

\end{table}

\begin{table}
\caption{\label{tab}} Elastic properties of the In-In impurity pair in GaAs. The columns give, respectively: (i) The relative position In-In positions (see Table I); (ii) The interaction energy; (iii) Bond length deviations induced by one of the impurities.

\bigskip

\begin{tabular}{|c|c|c|}
\hline
${\bf r}\left(\times \frac{a}{2}\right)$ & $\Delta E $ (meV) & ${\Delta L}$ ($10^{-3}$\AA ) \\ \hline
(110) & 6.194 & -2.776 \\ \hline
(220) & 6.665 & -1.407 \\ \hline
(330) & 1.952 & -0.430 \\ \hline
(002) & -4.077 & 0.713 \\ \hline
(004) & -0.637 & 0.072 \\ \hline
(211) & -2.869 & 0.468 \\ \hline
(222) & -0.939 & 0.128 \\ \hline
(321) & 0.447 & -0.140 \\ \hline

\end{tabular}
\end{table}

\begin{table}
\caption{\label{segreg}} Single In impurity energy as a function of depth for a free-surface supercell. Here, $n$ indicates the monolayer where the impurity is: $n=0$ is the free (top) monolayer, $n=1$ is the first monolayer below it and so on. The impurity energy increases with the distance from the free surface, meaning an elastic-induced effective attraction between impurity and surface. The energy converges to the bulk value at the third monolayer below the surface.

\bigskip

\begin{tabular}{|c|c c c c|}
\hline
$n$ & 0& 1 & 2 & 3 \\ \hline
Energy ($meV$)  & ~22  & 105 & 135 & 138 \\ \hline
\end{tabular}
\end{table}

\vfill\eject

\end{document}